# Sub-surface H$_2$S detection by a Surface Acoustic Wave passive wireless sensor interrogated with a ground penetrating radar

David Rabus,[1] Jean-Michel Friedt,[2] Lilia Arapan,[1] Simon Lamare[2,†] Marc Baqué,[3] Grégoire Audouin,[3] and Frédéric Chérioux[2*]

1 SENSeOR, 18 rue Alain Savary, F-25000 Besancon, France.
2 Univ. Bourgogne Franche-Comté, FEMTO-ST, CNRS, UFC, 15B avenue des Montboucons, F-25030 Besancon, France.
3 TOTAL SA, 2 place Jean Millier - La Défense 6, F-92078 Paris Cedex, France.

**ABSTRACT:** Long-term monitoring of organic pollutants in the soil is a major environmental challenge. We propose to meet this issue by the development of a polymer dedicated to selectively react with H$_2$S, coating surface acoustic wave transducers designed as passive cooperative targets with the compound, and probing their response using Ground Penetrating RADAR, thus providing the capability to monitor the presence of H$_2$S in the subsurface environment. The selectivity is brought by including lead(II) cation in a reticulated polymer matrix which can be deposited as a thin layer on a surface acoustic wave sensor. We demonstrate a signal enhancement mechanism in which water absorption magnifies the signal detection, making the sensor most sensitive to H$_2$S in an underground environment saturated with moisture.

*Keywords: Surface Acoustic Waves; Hydrogen Sulfide; Ground Penetrating RADAR; Wireless Measurement; Functional Polymer.*

Industrial infrastructures or human activities lead in most cases to the pollution of the corresponding soils. Therefore, land restoration after its use is one of the major environmental challenges in many countries for the development of a sustainable society. [1] For instance, the detection of hydrogen sulfide (H$_2$S) in the soil underneath the industrial infrastructures or in some buried industrial infrastructures, like pipelines, is of great importance because H$_2$S is a byproduct of many industrial activities [2] which is lethal at concentrations as low as 500 ppm. [3] Many works have demonstrated the development of sensors with high selectivity, low detection limit and fast response, in order to monitor the presence of H$_2$S in gas-phase. [4] The pollution of soils is mainly investigated by the classical drilling-treatment-analysis method. However, this method is expensive, slow and marred by errors. Therefore, development of powerful systems, aiming at monitoring the pollution in soils in order to replace the drilling-treatment-analysis, is required. [5] Underground sensing systems could emerge as an efficient solution if they were (i) easy to operate, (ii) energy efficient, (iii) accurate and reliable for long-term with referencing capabilities, and (iv) sensitive to preemptive warnings of geo- and environmental hazards. [3] These milestones require disruptive technological solutions to improve both the sensors and their interrogating systems, in order to unambiguously monitor *in-situ* the presence of organic pollutants in the subsurface, even over several decades. Within this objective, surface acoustic waves (SAW) sensors have been identified during the last two decades as one of the most promising passive embedded sensor platforms to deliver temporally and spatially on-demand sensing capable for measuring an event signal over a very long period (even years) thanks to cumulative measurement capability with no hindrance from a sensing element local energy source finite lifetime. Indeed, while SAW sensors have been initially designed as radiofrequency analog processing components, [6] they have been widely used for direct detection chemical sensors [7] and as passive cooperative target for short range RADAR sensing systems. [8] Nevertheless, only few investigations use the ability of SAW chemical sensors to be probed through a wireless link. [9] Another challenge then concerns the development of an interrogating system able to monitor the signal backscattered by the acoustic transducer through the soil. As opposed to short range wireless measurement techniques such as RFID, magneto-acoustic or magneto-elastic measurements, [10] or even inductively coupled short range interrogation we focus on long range propagative electromagnetic measurement techniques as implemented in RAdiofrequency Detection And Ranging (RADAR). Ground Penetrating RADAR (GPR) is a classical geophysical tool which is widely used to observe sub-surface interfaces, voids, leakages or buried objects (landmines, pipes etc.) associated with electrical conductivity variations. [11] However, up to now, the stability of some of GPR instruments time base prevents the fine measurement of the delay of the signal returned by the target (*i. e.* echo), mandatory stability of using GPR as interrogation system of a chemical SAW sensor. [12]

In this paper, we develop the first complete system allowing the detection of $H_2S$ in subsurface environments, *e.g.* in the context of leak detection, with no claim for quantitative concentration measurement. To achieve this objective, we develop 1) a SAW sensor acting as passive cooperative target and the associated sensitive layer compatible with the transducer to detect the $H_2S$ and 2) a Ground Penetrating Radar (GPR), playing the role of the interrogating system of the SAW sensor and operating from the surface controlled by a dedicated software for extracting the sensor signature and hence the detection of $H_2S$ in soils, even at a depth of 1 meter.

**Results**

*GPR and SAW sensor design*

Surface acoustic wave transducers meet both requirements of sensitive gravimetric sensors as well-known from the field of biosensing [7c] and wireless interrogation capability. [8] Indeed, they do not require any local energy source close to the sensing elements. Unlike the RadioFrequency Identification (RFID) tag communicating with a reader by using impedance modulated digital sentences, and whose measurement capability is hindered by the burden of large analog to digital conversion power consumption and the need to reach the rectifier diode voltage threshold, limiting its measurements range, the SAW transducer is based on the linear inverse piezoelectric conversion of the incoming electromagnetic wave collected by an antenna connected to the interdigitated transducer (IDT) electrode bus in order to generate an acoustic wave (Figure 1). All electrodes are patterned by lift-off using standard clean room facilities including photolithography of a thin layer of 250 nm evaporated aluminum. This acoustic wave propagates on the single crystal piezoelectric substrate with a velocity dependent on the boundary conditions including adlayers of molecules coating the propagation path. Ideally, a reference path with no chemical sensitivity allows for compensating for correlated effects (range from RADAR to sensor, temperature) while a measurement path acoustic velocity is dependent on correlated effects as well as chemical compound concentration (Figure 1). In our implementation, the whole sensor chip is coated with the chemical sensing layer and selective patterning of the reference path is still under investigation. This implementation allows for rejecting RADAR to sensor range nevertheless by analyzing delay differences between two echoes. After reflecting this wave using Bragg mirrors patterned on the piezoelectric substrate, the linear direct piezoelectric effect converts the acoustic wave back to an electromagnetic wave: from a user perspective, the SAW sensor is a radiofrequency electrical dipole returning echoes delayed by a duration dependent on a chemical compound layer thickness adsorbed on the sensor surface. Therefore, the objective of the design of the SAW sensor is mainly to optimize the returned power to extract the signal from noise by the readout system. The optimization of piezoelectric substrate electromechanical coupling coefficient and the lengthening of the delays of the returned echoes beyond clutter are the two design strategies allowing to reach interrogation ranges of multiple meters in low-loss subsurface environments. [13] In this work, as we aim at measuring ppm-range delay variations, we used a RADAR provided by Sensors and Software's SPIDAR control unit powering 100 and 200 MHz unshielded antennas because this apparatus possesses a very good stability of reference timescale clocking. We have indeed observed, on a one-hour long measurement, a standard deviation of 10 ps on a 0.3-microsecond echo delay difference, or a relative stability of 30 ppm at 2-second integration time falling to 10 ppm for a 1-minute integration time.

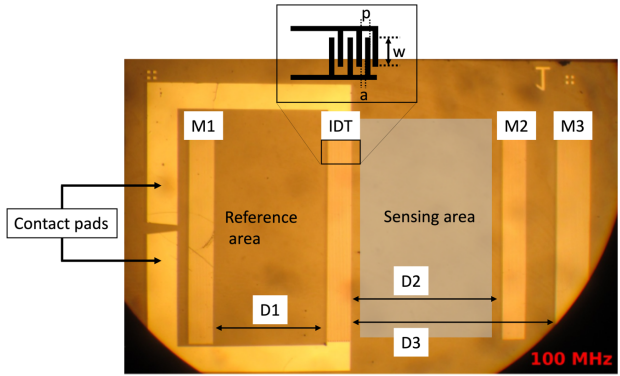

**Figure 1.** SAW sensor layout. The interdigitated transducer (IDT, with w = 3000 μm, a = 2p and p = 20 μm or 10 μm for a sensor working at 100 or 200 MHz, respectively) is located between three Bragg mirrors (M1, M2 and M3) positioned at different distances (D1, D2, D3) from the transducer in order to introduce different delays T. The contact pads are used for the electrical activation of IDT. M1 acts for the reference measurement. The active area is used for the deposition of the sensitive layer, M2 and M3 being used for the detection measurements.

The SAW sensor design includes an IDT, three mirrors and two contact pads. (Figure 1 and supporting information) The contact pads connected to the antenna are used for the electrical excitation of acoustic waves since the IDT induces an electric field in the piezoelectric substrate. Two parameters limit the possible values of the distances D between the three mirrors and IDT. D is given by D = T·c, with c the acoustic wave velocity dependent on the boundary conditions and hence the mass loading of the sensing layer and T the time delay. The lower value of D originates from the clutter suppression achieved by delaying the sensor echoes beyond all subsurface reflections assuming that the GPR receiver sensitivity only allows for recording signals reflected by interfaces located in soil at depths shallower than 100-m. This assumption assesses sensor echo time delay longer than one microsecond, which corresponds to D bigger than 1.5 mm, due to the typical soil permittivity around 2.25 or an electromagnetic velocity around 200 m/μs in soil and typical SAW velocities in the 3000 m/s range. The upper value for D is given by the acoustic propagation losses and sensor compactness which assesses D to be smaller than 4.5 mm (*i. e.* T below 3 microseconds). From the point of view of signal to noise ratio, the GPR is able to detect SAW sensor echo phase variations of at least few degrees. Indeed, classical time delay analysis using a spectral measurement, e.g. using a network analyzer, measures a phase rather than a time delay. If the system operates at 100 MHz, the SAW sensor acoustic path of 1.5 to 4.5 mm introduces a phase of $3.6·10^4$ to $1·10^5$ degrees. In a wireless measurement configuration using a pulsed RADAR, the phase measurement is either performed through the Fourier transform of the returned echoes or the cross-correlation of the echoes. [13] Considering the signal to noise ratio of the GPR acquisition allows for 10-ppm phase variation measurements, only phase changes of a few degrees can be accurately measured.

A key-point consists in the suppression of the delay introduced by the path between the reader unit and the subsurface sensor. This is successfully achieved by using differential measurements: since an echo delay is the sum of the time needed for the wave to travel from the GPR to the sensor to which the varying acoustic delay is added, a differential measurement cancels the former to only yield a time delay difference $T_{21}=2·(D2-D1)/c$. [14] Therefore, instead of only using Bragg Mirrors (M2 and M3) to return the echo from the active zone, another one (M1 in Figure 1) is deposited out of the chemical sensing area. This type of differential experiment requires an instrument with sampling rate or local oscillator stability better than these variations induced by the measurement (*i. e.* few ppm).

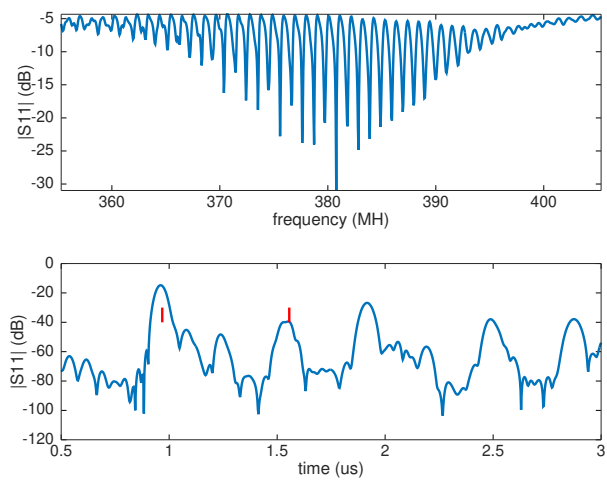

**Figure 2.** Frequency domain (top) and time domain (bottom) characterization of the reflective acoustic delay line used in the experiment depicted in this article. The red vertical markers on the time domain characteristics guide the eye to the echo delays analyzed, at 1.0 and 1.55 μs.

In our experiments, D1 = 2.0 mm and D2 = 3.0 mm yielding delays of 1.0 and 1.5 μs, respectively, with a center frequency selected to the GPR operating frequency as shown in Figure 2.

*Active sensitive layer*

We aim at functionalizing the SAW reflective delay lines with a compound selective and sensitive to hydrogen sulfide. The classical approaches of functionalization of SAW devices are based on the deposition of a thin film or the grafting of a monolayer of molecules onto a piezoelectric substrate. [15] Then, the exposure of SAW device to the targeted gas molecules lead to an ab- or adsorption (without any covalent link between analyte and the sensing films), on the sensitive layer. This ab- or adsorption promotes an increase of the film mass. This mass loading effect decreases the acoustic velocity and hence increases the time delay for a fixed path length, also observed as a phase decrease at a fixed delay. We aim at developing a tailored organic resist which 1) reacts selectively and irreversibly with $H_2S$, 2) is compatible with the classical processes of microfabrication used in clean-room, 3) yields irreversible reaction processes for cumulative measurements compatible with periodic probing of the sensor response as opposed to continuous monitoring.

The sulfur atom in $H_2S$ molecule exhibits a unique reactivity with metal ions, like $Pb^{2+}$ or $Zn^{2+}$, which inspired the well-known lead(II) acetate test paper. [16] Therefore, we consider introducing lead(II) acetate functional groups in a resist formulation, in order to be consistent with classical cleanroom spin coating for batch sensor fabrication. As many resists used in micro- nanofabrication processes are based on the thermo- or photopolymerization of epoxides with nucleophiles (like alcohol or amine derivatives), we have first synthesized two α,ω-dihydroxyalkyl lead(II)carboxylate complexes which are ended by two hydroxyl functions (see Scheme 1 and supporting information for experimental details).

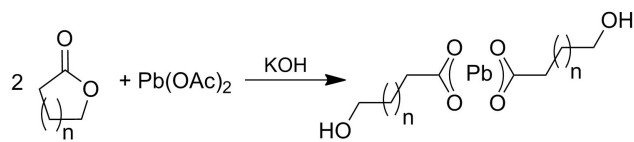

**Scheme 1**. Synthesis of α,ω-dihydroxyalkyl lead(II) carboxylate complexes with n = 1 or 2.

This reaction is versatile and efficient. The ring-opening of a lactone in presence of a strong base and lead(II) nitrate gives quantitatively the expected lead(II) carboxylate ended by the OH functions. The length of the alkyl chains in final complexes is fixed by the choice of the starting lactone.

Then, in order to achieve the formation of the resist, we have mixed the tailored α,ω-dihydroxyalkyl lead(II)carboxylate complexes (as monomer, 7 mmol) with 1,3-butadiene diepoxide as co-monomer (14 mmol), 1,3,5-trihydroxy benzene as reticulating agent (5 mmol) in a DMF solution (74 mmol).

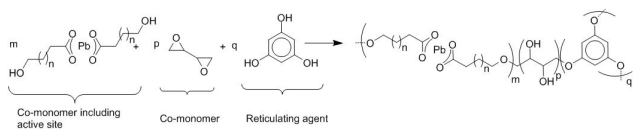

**Scheme 2.** Synthesis of the reticulated copolymer including lead(II) carboxylate sites as active centers for the sensing of $H_2S$.

As the final co-polymer is expected to be highly reticulated, and consequently quite insoluble, we have directly deposited the mixture on the SAW devices. To achieve this functionalization, a drop of this solution was spin-coated (acceleration: 1000 rpm/s, speed: 2000 rpm, duration: 10s) onto a wafer containing the SAW reflective delay lines. Finally, the film is annealed at 180°C for 10 min. This procedure leads to the formation of a solid thin film (thickness: 250 ±20 nm, measured by mechanical profiler) which covers entirely the wafer (see Figure S1 in supporting information). This simple and reproducible procedure is fully compatible with a collective fabrication process at the wafer scale, although selective patterning of the sensing area and leaving the reference area insensitive to chemical reaction is still under investigation, with a lift-off process having been demonstrated but requiring manual removal of the sacrificial resist layer.

*Demonstration of the sensing-capability of the SAW-functionalized components*

In a first experiment, we aim at demonstrating the capability of the tailored resist to detect efficiently $H_2S$ in a gas phase. A typical experiment is based on the exposure of a SAW sensor fully coated with the sensing layer to a flux of $H_2S$ and/or $H_2O$, each one being in gas phase. The measurement is performed initially in a wired configuration, with a Rohde & Schwarz ZVC8 vector network analyzer, with the time domain measurement option collecting the frequency domain reflection coefficient and performing the inverse Fourier transform to provide the returned power and phase at the time delay of the various echoes. All following plots exhibit the phase difference between echoes as a function of measurement time.

Three experiments have been planned to investigate the sensing capability. The first one consists in the exposure of the SAW sensor (operating at 380 MHz) to a dry $H_2S$ atmosphere at room temperature.

As soon as the SAW sensor is exposed to the dry atmosphere of $H_2S$ (20% in dry $N_2$), we observed a strong decrease of the phase (Figure 3).

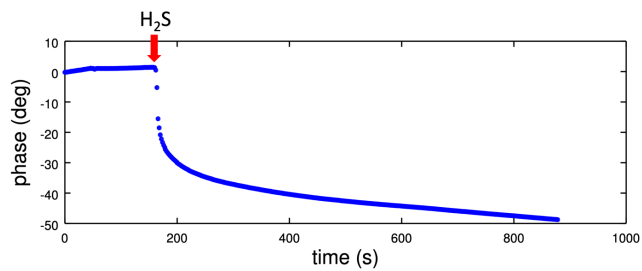

**Figure 3.** Exposure of a SAW sensor to a dry atmosphere of $H_2S$.

The phase variation, dPhi, is 50° after 800s, corresponding to a sensitivity of 700 ppm (Figure 3). This phase variation of 50° can be assigned to a mass loading during the exposition of the SAW sensor. On the one hand, the gravimetric sensitivity (S=(dPhi/Phi)·(A/$D_m$) with dPhi the phase variation around the phase rotation along the path Phi induced by a mass $D_m$ variation over area A) of a Rayleigh wave is assumed to be in the S=100 cm²/g range. [17] This definition of sensitivity handling surface mass density and relative phase (or frequency) variation because it cancels the actual path length and is valid both for transmission and reflective delay lines. Considering a delay of t = 1.55 μs and an operating frequency of f= 380 MHz, the phase rotation along the path is Phi=360·t·f = 212040 degrees. The corresponding mass variation $D_m$ per unit area is $D_m$/A = dPhi/(S·Phi) where dPhi is the phase variation (*i. e.* 50°, Figure 3) and the area A is 0.3·0.3 cm² (Figure 1). The numerical application hints a mass variation of about 2.4 μg/cm², since the resist thickness is small enough to only act as a perturbative layer, or a hydrogen sulfide mass increment of 0.21 μg in the layer. On the other hand, the expected absorbed mass of hydrogen sulfide is estimated as 0.16 μg. Indeed, the molecular weight of the polymer is 692.6 g/mol (Scheme 2). Assuming a density of 1.4 g/cm³ for this polymer, [18] the volume of sensitive layer of the SAW sensor is 250·10⁻⁷·0.3·0.3 cm³, which correspond to 4.55 nmol. By considering the molecular weight of 34 g/mol for $H_2S$, the layer loads an added mass of 0.16 µg. This value is very close to the expected 0.21 µg deduced from the acoustic signal analysis assuming a purely gravimetric effect. We conclude that in a dry saturated atmosphere, assuming all lead sensing sites react with hydrogen sulfide, that mass loading is sufficient to interpret the data.

Another set of experiment is made by using an air atmosphere saturated with water moisture at ambient pressure at 20°C and, then exposed to an atmosphere containing 20% of $H_2S$ and water saturated pressure at room temperature (Figure 4).

When the SAW sensor (operating at 380 MHz) is exposed to an atmosphere saturated with $H_2O$, the phase variation is 12° (inset in Figure 4). Then, the exposure to $H_2S$ leads to an impressive phase variation of over 650° (or 9000 ppm), which is 13 times stronger than those observed in a dry atmosphere (Figure 3).

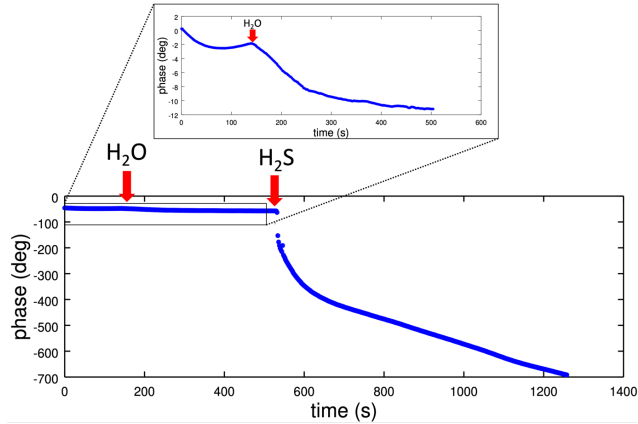

**Figure 4.** Exposure of a SAW sensor to atmosphere saturated with $H_2O$, then followed by $H_2S$ (20%) in a dry atmosphere. The phase variation due to the $H_2O$ is of 12° (inset) while the exposure to $H_2S$ leads to a phase variation of 650°.

Finally, another experiment is performed by only inverting the exposure of a SAW sensor to $H_2S$ then $H_2O$ (Figure 5) by comparison with the previous experiment.

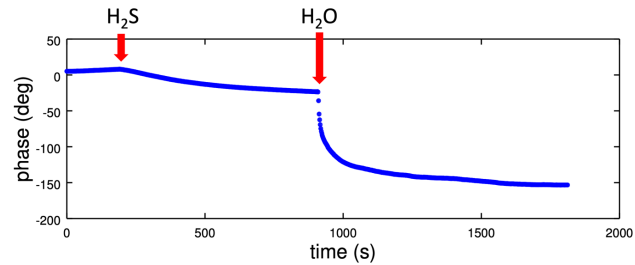

**Figure 5.** Exposure of a SAW sensor to atmosphere saturated with $H_2S$ (20% in a dry atmosphere), then followed by atmosphere saturated with $H_2O$. The phase variation due to the $H_2S$ is 40° while the exposure to $H_2O$ leads to a phase variation of 100°.

After exposure with dry $H_2S$, the phase variation is of 40°, which is close to those observed in the first experiment (Figure 3). However, as soon as the SAW sensor is exposed to a water-saturated atmosphere, the phase variation is 100° (Figure 5). The analysis upon water loading is much more complex since the mass loading is accompanied with polymer structure change which might affect its Young modulus and hence slow down the wave upon softening. [19]

*Subsurface sensor measurement*

The previous experiments demonstrated the capability of the SAW sensor to detect $H_2S$ in a gas phase with wired configuration and in a controlled environment. We aim at investigating the efficiency of this SAW sensor to detect $H_2S$ in a wireless configuration by using a GPR as interrogation system. To achieve this experiment, a wireless reflective delay line SAW sensor is inserted in polyvinyl chloride (PVC) pipes which are buried at different depths ranging from 10 to 100 cm in a sandbox (Figure S2). [12] The sandbox is located outside the building and no specific condition (pression, temperature, weather etc.) was defined in order to simulate a "real condition of use". The sensor is probed from the surface by recording once every 2-seconds a GPR A-scan collected by the Sensors and Software SPIDAR controller connected to PulseEkko emitter and transmitted unshielded 100 or 200 MHz antennas.

The sensor is initially inserted in the PVC pipe, and a baseline is reached after a few minutes as measured continuously by the GPR custom control software. Once the stable baseline is reached, $H_2S$ is generated *in situ* by the reaction of an aqueous solution of HCl acid with a FeS powder, leading to the formation of $H_2S$ and $FeCl_3$. The $H_2S$ produced gas is pushed by a fan in the PVC tube in order to expose the SAW sensor to this gas.

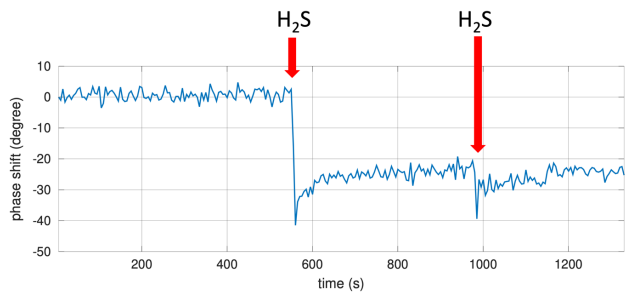

**Figure 6.** 100-MHz sensor located 1-m deep in sand: $H_2S$ is generated at times 570 s and 1000 s. The first exposure to $H_2S$ leads to a phase variation of 25°. The second exposure does not induce a phase variation indicating that the reactive sites are saturated upon the initial exposure.

The exposure of the SAW sensor leads to a phase variation of 25° which is easily detected by the GPR through one-meter thick sand (Figure 6). Additional exposure does not lead to a phase variation with these experimental conditions. The repeatability of wireless measurements is checked by using two different 200-MHz sensors located at different depths (0.3 and 0.4 m, see Figures S3 and S4, respectively). The observed phase variations are 130 and 220°, respectively. These three experiments demonstrate that the combination developed in this work based on a tailored sensitive resist, a SAW sensor and a GPR, is efficient to detect $H_2S$ in a gas phase in a wireless configuration in soil and in real conditions (weather, temperature, humidity, etc.).

**Discussion**

On the basis of all experimental data, we demonstrate that the complete system described in this article is efficient to detect in sub-surface the presence of $H_2S$ in a gas phase. All phase variations observed are significant (above 25° phase shift or 2000 ppm) with respect to the GPR noise level (6° phase fluctuation) thanks to the mass loading enhancement induced by the presence of moisture, with phase variations consistent with those found in the literature using other mechanisms. Indeed, 3 kHz frequency variation of a 60 MHz SAW device with a gap between IDTs of 2.5 mm, leading to an equivalent phase shift of 900 degrees was reported. [20] The origin of this phenomenon is explained as follows: the exposure of a lead(II) acetate derivative to $H_2S$ leads to the formation of PbS and carboxylic derivative (Figure 7). The formation of PbS explains the color variation in the well-known lead(II) acetate test paper [16,21].

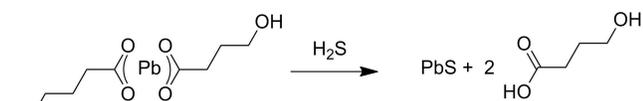

**Figure 7.** Reaction of $H_2S$ with a molecular lead(II) acetate derivative.

In addition, as water has no role in this polymer matrix breaking process, the presence or the absence of water (or moisture) alone is hardly detected by the sensor as observed in Figure 4 before injection of $H_2S$. However, when a sensor is exposed to $H_2S$ followed by $H_2O$ or $H_2O$ followed by $H_2S$, we observe a strong enhancement of the resulting phase variation, respectively 600° and 100°. The enhancement mechanism is explained by the depolymerization of the resist when exposed to $H_2S$. Indeed, the starting resist is initially compact (due to the reticulation) and hydrophobic, which prevents interaction with water (moisture). However, upon reaction with $H_2S$, the lead(II) cations which are involved in the polymer chains are cleaved to produce PbS clusters and short oligomers ended by several carboxylic moieties (Figure 8). These carboxylic groups are hydrophilic due to the formation of multiple hydrogen bonds with water. As the skeleton of polymers is strongly altered and now turns to be strongly hydrophilic, water molecules are absorbed in the organic layer, which acts as a sponge, leading to a significant mass loading enhancement and hence slower acoustic wave, inducing the observed phase variations.

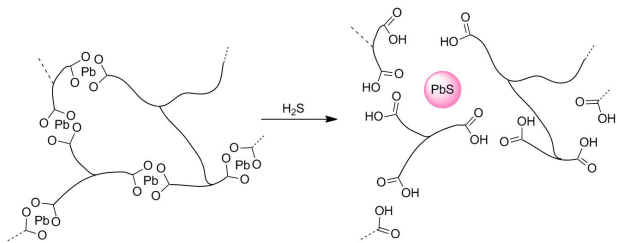

**Figure 8.** Depolymerisation of copolymer leading to the creation of hydrophilic functions (carboxylic acid) and lead sulfide cluster due to the exposure to $H_2S$.

This important phase enhancement is useful for the wireless detection in which signal to noise ratio is degraded by radiofrequency signal attenuation when propagating through soil (here sand) leading to poorer detection limits than in the wired configuration. Eventually, as expected from the irreversible reaction of $H_2S$ with the $Pb^{2+}$ cations, if the initial exposure of the SAW sensor induces complete transformation of all active sites, further exposure does not lead to additional phase variation (Figure 5, date 1000 s).

**Conclusion**

The combination of a tailored sensitive resist, a SAW sensor and a GPR, is used as an efficient system to detect $H_2S$ in a gas phase in a wireless configuration in soil and in real conditions (weather, temperature, humidity, etc.). The strong phase variation, required to compensate for radiofrequency signal attenuation when propagating through soil, is based on an enhancement of the mass loading due to the presence of water or moisture. This remarkable mass loading is justified by the initial design of the resist which was selected in order to induce its depolymerization during the exposure to $H_2S$. As this reaction is irreversible and the sensor is battery-less, it is an ideal candidate for long-term cumulative detection of the presence of $H_2S$ in subsurface: phase shifts accumulate as pollutants flow over the sensor and even if sensor monitoring is periodic and not continuous, a punctual pollutant spike is recorded and will be measured during the next sensor probing session. This new strategy paves the way of the developments of sensors, and their interrogation by GPR, for land preservation of industrial sites.

**AUTHOR INFORMATION**

**Corresponding Author**

frederic.cherioux@femto-st.fr

**Present Addresses**

† Simon LAMARE, e-mail: simon.lamare.pro@hotmail.fr
BioParc, 850 bld Sebastien Brant, 67400 Illkirch

**Author Contributions**

All authors contributed equally.

**Funding Sources**

No competing financial interests have been declared.

**ACKNOWLEDGMENT**

This work is supported by the French National Agency through the project UNDERGROUND (ANR-17-CE24-0037). The authors also acknowledge partial support of the French RENATECH network and its FEMTO-ST technological facility. The authors thank S. Benchabane, L. Morgenthaler and P. Tanguy (FEMTO-ST), F. Gégot (SENSeOR), J. R. Ordonez-Varela, A. Credoz, A. Le Beulze and M.-F. Bennassy (TOTAL) for fruitful discussions.

*Mater.* **1996**, *8*, 1919-1924. c) Zeng, Z.; Wang S.; Yang, S. Synthesis and Characterization of PbS Nanocrystallites in Random Copolymer Ionomers. *Chem. Mater.* **1999**, *11*, 3365-3369.

**Synthesis of complexes**
53 mmol of γ-butyrolactone (or ε-caprolactone) and 53 mmol of KOH are dissolved in 150ml of ethanol and 30 ml of $H_2O$. The mixture is heated at reflux for 1h. Then, 26 mmol of $Pb(NO_3)_2$ in 40 ml of $H_2O$ are added. Then, the mixture is heated at reflux for 20 min. The solvents were removed under reduced pressure. The crude solid is washed in 50 ml of hot ethanol, to give the expected complexes.

*Lead(II) 4-hydroxy-butanoate*
$^1$H NMR (300 MHz, $D_2O$) δ = 3.50 (t, J=7.4Hz, 4H), 2.14 (t, J=7.4Hz, 4H), 1.70 (q, J=7.4HZ, 4H). $^{13}$C NMR (75 MHz, $D_2O$) δ = 183.1 (C=O), 61.6 (C-OH), 34.1, 28.4.

*Lead(II) 6-hydroxy-hexanoate*
$^1$H NMR (300 MHz, $D_2O$) δ = 3.57 (t, J=6.5Hz, 4H), 2.19 (t, J=7.40Hz, 4H), 1.66-1.55 (m, 8H), 1.66-1.55-1.50 (m, 4H). $^{13}$C NMR (75 MHz, $D_2O$) δ = 184.1 (C=O), 61.5 (C-OH), 38.3, 31.1, 25.0, 24.9.

**SAW device fabrication**
Ground Penetrating RADAR operating frequency is a tradeoff between antenna dimensions and interrogation range: reaching the 1-3 m depth targeted to reach the water table or the sub-surface infrastructure requires operating in the 100 to 400 MHz range. Thus, YXl/128° lithium niobate acoustic delay lines were manufactured. With a wavelength of 40 μm and a metal thickness of 2.5% wavelength, the L = 2 to 3 mm long acoustic path introduces a two way trip phase rotation of 2xLx360xf/c =36180 to 54270° phase rotation or a differential phase accumulation of 18090° considering the Rayleigh wave velocity c = 3980 m/s and an operating frequency of f = 100 MHz, or a differential phase rotation of 72360° for the same geometry of a sensor operating at 400 MHz. In order to avoid damping the Rayleigh wave, the organic sensing layer will be designed as thin with respect to the wavelength, aimed at varying the acoustic velocity without excessive acoustic losses: a thickness of about 1% wavelength is considered, yielding 100 to 500 nm thick adlayers for devices operating in the 100 to 400 MHz range.

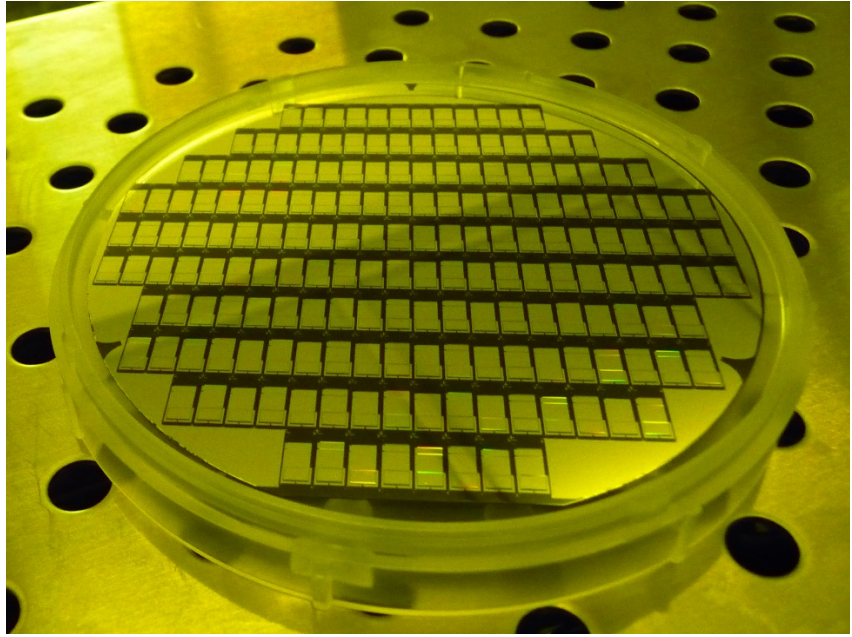

**Figure S1.** A 4" wafer composed of 200 SAW components completely covers by a solid layer (thickness: 250nm) of the tailored-resist containing the Pb(II)-site.

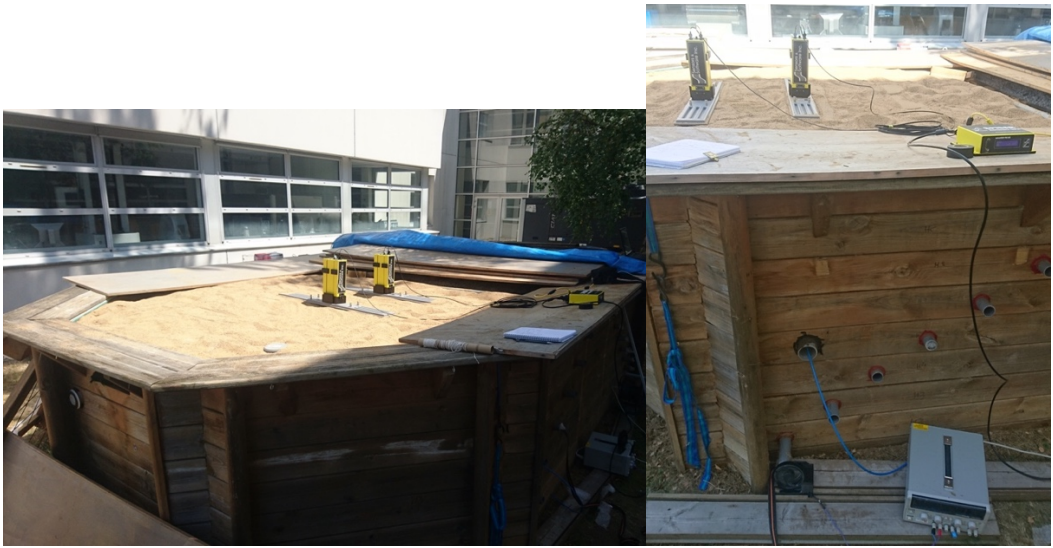

**Figure S2.** The sand-pit for the demonstration of sub-surface wireless sensing (left). The GPR is over the sand-pit and the sensor is located in a pipe where $H_2S$ is injected into the pipe as gas phase (right) at different depths (from 0.2 to 1m).

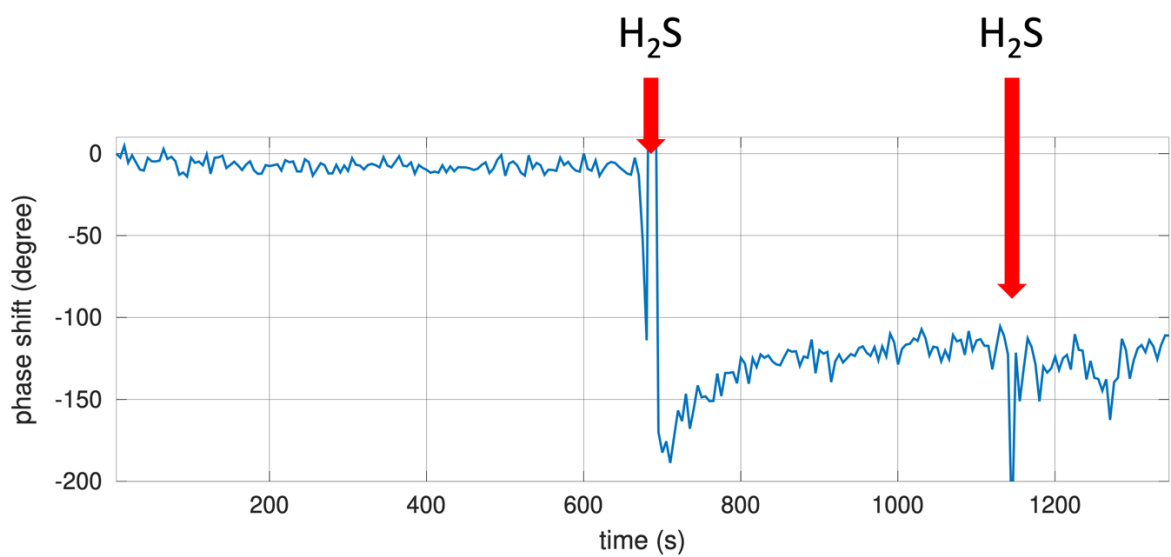

**Figure S3.** 200-MHz sensor located 0.3-m deep in sand: H2S is generated at times 695 s and 1150 s. The lack of irreversible phase variation upon the second exposure indicates that the reactive sites were saturated upon the initial exposure.

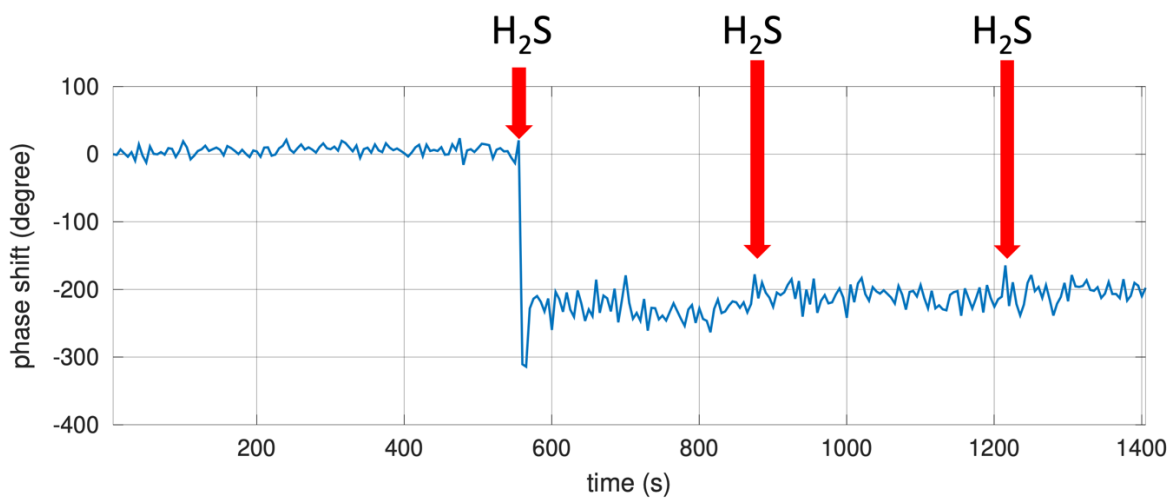

**Figure S4.** 200-MHz sensor located 0.4-m deep in sand: H2S is generated at times 565 s, 875 and 1235 s. The lack of irreversible phase variation upon the second and third exposures indicates that the reactive sites were saturated upon the initial exposure.